\documentstyle[aps,pre,amssymb,graphicx]{revtex}

\begin{document}
\draft
\title{NATURE OF SONOLUMINESCENCE: NOBLE GAS RADIATION EXCITED BY HOT ELECTRONS IN
"COLD" WATER}
\author{N.Garc\'\i a$^{1}$, A.P.Levanyuk$^{2}$ and V.V.Osipov$^{1,3}$}
\address{$^1$Laboratorio de F\'\i sica de Sistemas Peque\~nos y Nanotecnolog\'\i a,\\
Consejo Superior de Investigaci\'{o}nes Cient\'\i ficas, c/Serrano
144, 28006 Madrid, Spain}
\address{$^{2}$Departamento de F\'\i sica de la Materia Condensada, \\
Universidad  Aut\'{o}noma de Madrid, 28049 Madrid, Spain}
\address{$^{3}$Department of Theoretical Physics, Russian Science Center
"ORION", \\ Plekhanova str. 2/46, 111123 Moscow, Russia}
\date{\today}
\maketitle

\begin{abstract}
It was proposed before that single bubble sonoluminescence (SBSL) may be
caused by strong electric fields occurring in water near the surface of
collapsing gas bubbles because of the flexoelectric effect involving
polarization resulting from a gradient of pressure. Here we show that these
fields can indeed provoke dynamic electric breakdown in a micron-size region
near the bubble and consider the scenario of the\ SBSL. The scenario is: (i)
at the last stage of incomplete collapse of the bubble, the gradient of
pressure in water near the bubble surface has such a value and a sign that
the electric field arising from the flexoelectric effect exceeds the
threshold field of the dynamic electrical breakdown of water and is directed
to the bubble center; (ii) mobile electrons are generated because of thermal
ionization of water molecules near the bubble surface; (iii) these electrons
are accelerated in ''cold'' water by the strong electric fields; (iv) these
hot electrons transfer noble gas atoms dissolved in water to high-energy
excited states and optical transitions between these states produce SBSL UV
flashes in the trasparency window of water; (v) the breakdown can be
repeated several times and the power and duration of the UV flash are
determined by the multiplicity of the breakdowns. The SBSL spectrum is found
to resemble a black-body spectrum where temperature is given by the
effective temperature of the hot electrons. The pulse energy and some other
characteristics of the SBSL are found to be in agreement with the
experimental data when realistic estimations are made.
\end{abstract}


%%%%% end of frontmatter %%%%%%%%%

%%%% BODY TEXT %%%%%%%%%%%

\section{Introduction}

Sonoluminescence refers to the phenomenon of light emission during acoustic
radiation of a liquid and is associated with cavitation bubbles present in
the liquid. The most controllable and promising experimental data were
obtained for single bubble sonoluminescence (SBSL): picosecond UV radiation
of a single bubble pulsating in the field of the sound wave \cite
{gait,review}.

Though the case of a single pulsating bubble is, of course, much simpler
than cavitation in general, it turns out that the SBSL is a very complex
phenomenon which still remains not completely understood. There is a vast
literature devoted to SBSL that was recently reviewed in Ref.\cite{review}.
Interest in this phenomenon is stimulated, on the one hand, by the fact that
in relatively simple and controllable experiments extraordinary conditions
(ultra-high pressures, temperatures, and ultra-short light flashes) are
realized at the final stage of the bubble collapse; on the other hand, it
looks promising to use the SBSL to construct a source of ultraviolet
ultra-short flashes that is much cheaper than lasers.

The highly involved hydrodynamics of bubble collapse has been addressed in
many papers that are reviewed in Ref. \cite{review}. One of the most
important problems to explain here is the very existence of stable pulsation
of bubbles. Recently a theory has been developed \cite{LohsePRL97,LohseJCP97}
which explains this regime in terms of the dissolved gas diffusion and
chemical reactions in the gas within the bubble. It has been shown that an
accumulation of noble gas in the bubble takes place during pulsation of the
bubble and that in a stable situation the gas in the bubble consists, almost
entirely, of noble gas and, of course, of water vapor. These theoretical
conclusions were supported by the experimental data of Ref.\cite{Matula}.

Another key problem in SBSL is the mechanism of the light emission. Many
mechanisms have been proposed, criticized, and reviewed to explain UV
flashes radiated by the collapsing bubble \cite{review}. The most popular of
the suggested mechanisms are the adiabatic heating of the bubble gas \cite
{review,Lofs93,Hilg}, shock wave-Bremsstrahlung model \cite
{review,Jar60,Lofs92,Wu93}, and, recently proposed proton-tunneling
radiation as a result of a phase transition in water at ultra-high pressure
\cite{Willison}. A general feature of these mechanisms is that extraordinary
conditions are needed which can only be realized, if at all, at a very small
bubble radius when the density of the bubble gas is close to the water
density and both the hydrodynamics of the bubble and the properties of the
bubble content and the neighboring liquid are not known from experiment but
inferred from numerical simulations.

In this paper we show that the main characteristic features of SBSL can be
explained even without making any assumptions about the extraordinary
conditions. We are far from saying that these extraordinary conditions are
not present in experiments. What we mean is that there is another mechanism
of the SBSL which occurs in water near the bubble surface but not in the
bubble gas as has been assumed in the most popular models of SBSL \cite
{review}. Extraordinary conditions are not necessary for the operation of
this mechanism even though it may well be the main cause of UV radiation.
The mechanism under consideration is based on the idea put forward in Ref.%
\cite{GL} that SBSL occurs because of electric breakdown in strong electric
fields arising near the bubble surface as a result of flexoelectric effect,
that is the effect of polarization of water because of gradients of pressure
\cite{Tagan}. Here we present a scenario and estimations that show that
within the hypothesis the main features of SBSL can be explained using
relatively moderate parameters, e.g., temperatures (5$\div $10)10$^{3}$K for
the bubble gas \cite{Lofs93}, and the natural (expected) value of the
flexoelectric coefficient of water \cite{GL}.

Let us mention that Ref.\cite{GL} has left many questions to answer. The
origin of the optical radiation of the bubble remaines unexplained. The
problem is that in the visible region pure water has very low levels of
absorbtion and radiation due to interband optical transition \cite{Will74}.
Moreover the breakdown scenario is far from being clear. The mechanism of
the breakdown in water (see, e.g.,\cite{Sachi}) involves ''lucky electrons''
whose acceleration in the electric field leads to development of an
avalanche. At ambient temperature the concentration of free electrons in
water is quite negligible (the band gap is about 6.5 eV \cite{Will74,Sachi})
and it is impossible to find a ''lucky electron'' in a small volume near the
bubble surface, i.e. in the region of high electric field, during the short
time of the existence of this field. The spectacular synchronization of the
emission pulses \cite{review,synchr} that in the case of breakdown appears,
at least at first sight, to be hardly compatible with the fact that we are
dealing with a probabilistic situation. In addition, the reference to the
Penning effect to explain the role of the noble gases is not convincing
because the ionization energy of the metastable state is more than the band
gap of water as distinct from the case of breakdown in gases where the
Penning effect is pronounced.

In this paper we consider in more detail possible processes associated with
the electric field arising from the flexoelectric effect near a bubble
exhibiting SBSL. A common difficulty with theories of SBSL is that little is
known about the parameters of gas in the bubble when the bubble radius is
near its minimum value. Consensus exists about one point: it is far from
being a gas, because the lowest limit of the radius is governed by the van
der Waals repulsion and the minimum volume of the bubble is close to the van
der Waals hard core volume. The equations of state used for these conditions
are not reliable at present. This is a challenging and fundamental problem
but it is beyond the scope of this paper. What is now possible is to make
order-of-magnitude estimations. That is why we shall first explain what
values of relevant parameters are necessary for our scenario to be operative
and then discuss whether our assumptions about these parameters are
realistic. We include the value of the flexoelectric coefficient of water
among these parameters as well. This coefficient has not been measured,
unfortunately, and we use its estimated ''natural'' value but, from the
other hand, the knowledge of its precise value would hardly be of decisive
importance because the bubble gas parameters are not known precisely.

Let us describe shortly the proposed scenario. At certain short time
interval, $\tau _{c}\sim 1$ns, when the bubble radius is near its minimum,
the acceleration of the radius and, therefore, the pressure gradient, assume
gigantic values \cite{review} and the sign of the pressure gradient is such
as to create a strong (depolarizing) electric field directed to the centre
of the bubble in a thin water layer, $\sim 1\mu m$, near the bubble surface,
because of the flexoelctric effect. What happens then is, in effect,
screening of this depolarizing field by free electrons. Indeed, during the
same time interval the temperature of the gas sharply increases up to at
least several thousand K, which owing to thermal excitation of the bubble
gas and water in a thin layer near the bubble surface makes the
concentration of free electrons appreciable. The free electrons are
accelerated by the strong ''flexoelectric'' field up to energies sufficient
to generate additional free electrons as a result of the electric breakdown
of water. The hot electrons also collide with noble gas atoms dissolved in
water and transfer them to higher energy excited states. Optical transitions
from these states produce light radiation with a broad spectrum whose shape
is determined mainly by the energy distribution of the hot electrons. The
latter has the form of a Maxwell distribution with an effective electron
temperature which can be very high. This is due to the fact that the noble
gas atoms have a huge number of excited states with very high probabilities
of optical transitions between them (see, e.g., \cite{Baum62,Karlov}). Since
the radiation occurs in the region of very strong and inhomogeneous electric
fields the observed radiation spectrum is featureless. The characteristic
time of the electric breakdown is much less than the characteristic time for
the last stage of the collapse $\tau _{c}$, i.e. the polarization can
continue to change after the first breakdown and the electric field can
reach the breakdown threshold value more than once. As a result, several
breakdowns can take place during the time interval $\tau _{c}$. During each
breakdown the noble gas atoms are excited. An important specific feature of
the noble gas atoms is the existence of long-living (metastable) excited
states with a life time of up to several milliseconds (see, e.g., \cite
{Baum62,Karlov}). Therefore, once excited the noble gas atom can remain in
the metastable state for the entire time interval $\tau _{c}\sim 1$ns, and,
possibly, for many periods of the acoustic wave (the period is about $30\mu $%
s). That means that in the stationary state the longer is the pulse the more
is its intensity.

The paper is organized as follows. In Sec.2 we discuss the flexoelectric
effect in water in more detail than in \cite{GL} and consider the sign of
the flexoelectric field, which is crucial for the proposed scenario, at
different stages of the bubble collapse. The scenario of SBSL is presented
in Sec.3 where the values of relevant parameters necessary for realizing the
scenario are estimated. In Sec.4 we discuss the energy distribution
functions of hot electrons and the breakdown conditions in strong
flexoelectric fields as well as the spectrum of SBSL. In Conclusions we
summarize the results of our theory, discuss their relation to the
experimental data, speculate about some possible modifications of the
scenario and point out some problems to solve.

\section{Flexoelectric effect and electric field near the bubble}

We have already mentioned that the proposed mechanism of SBSL is based on
the flexoelectric effect, namely the appearance of a polarization (electric
field) due to gradient of density (pressure)\cite{GL}. This effect is not
widely known and it was discussed in \cite{GL} but shortly. Besides, the
flexoelectric coeficients are still not determined experimentally\cite{LC}.
That is why we think it is worthwhile considering in more detail the
appearance of electric field due to flexoelectric effect .

The flexoelectric effect is a particular case of a more general phenomenon:
appearance of electrical polarization, ${\bf P,}$as a result of gradient of
some scalar quantity, e.g., temperature, concentration of a component, mass
density, $\rho $. This effect should take place in any substance \cite{Tagan}%
. As usual, the material equation can be written in several forms. In
particular, considering polarization as a function of dilatation, $u=\Delta
\rho /\rho $, and the electric field, ${\bf E}$, one has for an isotropic
medium
\begin{equation}
{\bf P}{\bf =}\alpha \nabla u{\bf +\varepsilon }_{0}\chi {\bf E},
\label{polar}
\end{equation}
where $\chi $ is the electric susceptibility and $\alpha $ is one of the
flexoelectric coefficients. We shall be interested in the case of spherical
symmetry and the absence of free charges. In this case ${\bf D}=\varepsilon
_{0}{\bf E}+{\bf P}=0.$ Then from Eq.\ref{polar} it follows that

\begin{equation}
{\bf E}=-\frac{\alpha }{\varepsilon \varepsilon _{0}}\nabla u=-\frac{\alpha
\beta }{\varepsilon \varepsilon _{0}}\nabla p\equiv f\nabla p,  \label{El}
\end{equation}
since $u=\beta p$, where $p$ is the (excess) pressure, $\beta $ is the
compressibility.

To estimate the flexoelectric coefficients it is convenient to consider
separately two main mechanisms of polarization: charge displacement and
dipole ordering. We shall see that in both cases the coefficient of
proportionality between polarization and gradient of pressure, $f$, has the
same characteristic value.

The charge displacement polarization is realized, e.g., in ionic crystals.
When a gradient of dilatation takes place there are less and more compressed
regions in each unit cell. The ions of larger radius tend to displace to the
less compressed region while ions of less radius displace into the opposite
direction. Since cations and anions usually have different radii their
displacements produce polarization. The coefficient $\alpha $ can be
estimated as follows \cite{Tagan}. The maximum (''atomic'') gradient of
dilatation is equal to $1/d,$ where $d$ is the interatomic distance.This
gradient has to produce ''atomic'' polarization $P_{at}\sim e/d^{2}$, where $%
e$ is the electron charge, when the electric field is absent (compensated by
some charges), i.e. $\alpha _{ion}\sim e/d$ . Then from Eq.\ref{El} it
follows that

\begin{equation}
f\sim \frac{\alpha _{ion}\beta }{\varepsilon \varepsilon _{0}}\sim \frac{%
e\beta }{d\varepsilon _{0}}  \label{flexnonpol}
\end{equation}
where we have taken into account that $\varepsilon \sim 1$ for ionic
crystals.

We shall argue that the same estimation is aplicable for materials with
dipole ordering. The dipole ordering polarization in the absence of an
external electric field but under a gradient of pressure arises because of
geometric asymmetry of the molecular dipoles. For example, the geometric
shape of water molecule is highly asymmetric: the negatively charged end is
much more compact than the positive one: the negative charge is concentrated
in the oxygen ion (O$^{--}$) while the positive charge is shared by two
hydrogen ions (H$^{+}$) located fairly far from each other. Under a pressure
gradient a water molecule tries to orient itself in such a way that the
oxygen ion would be located in the region of higher density while the
hydrogens would be located in the region of lesser density. In other words,
a gradient of pressure leads to orientation of the water molecular dipoles.
It is well known that polarization due to dipole ordering is much more
effective than polarization due to charge displacements. This is reflected
in the fact that dielectric constants of dipolar materials are, normally,
much larger than of non-polar materials. For the same reason it is natural
to estimate $\alpha _{dip}\sim \varepsilon \alpha _{ion}$. Then from Eq.\ref
{El} it follows that for water

\begin{equation}
f\sim \frac{\alpha _{dip}\beta }{\varepsilon \varepsilon _{0}}\sim \frac{%
e\beta }{d\varepsilon _{0}}\sim 10^{-7}\frac{Vm^{2}}{N},
\label{flex}
\end{equation}
where we have taken into accout that for water $\beta \!\approx 5\cdot
10^{-10}\frac{m^{2}}{N},$ $d\sim 10^{-10}m$.

For what follows the sign of the flexoelectric coefficients is of
importance. As it has been mentioned above the negative tip of the water
molecule, the oxygen ion, tends to be located in the region of higher
pressure. This means that the polarization vector is directed opposite to
the gradient of pressure, i.e., the flexoelectric coefficient $\alpha $ is
negative and the coefficient $f$ is positive.

Let us mention that we rather underestimated the flexoelectric
coefficient of water than overestimated. Indeed, what we call
$\alpha _{ion}$ is replaced, in fact, by $\alpha _{at}$ defining
an ''atomic'' flexoelectric coefficient that is of the same order
of magnitude for all substances, i.e. it does not take into
account specific features of the substance in question. It is
natural to expect that a substance consisting of molecules whose
electric asymmetry (existence of a dipolar moment) is accompanied
by a pronounced geometric asymmetry (as it is for water) will
exhibit a stronger flexoelectric effect than that estimated above.
However, while this coefficient is not measured (reliable
calculations seem much more problematic than measurements) we
shall assume the value given by Eq.\ref {flex}. On the contrary,
the conclusion reached about the sign of the coefficient $f$,
which is crucial for the mechanism discussed in this paper, seems
much more definite.

Above we have neglected the conductivity of water. This may seem
questionable because the Debye radius of the electric field screening in
water is comparable with the characteristic size of the strong field region (%
$\sim 1\mu $m). However, this neglection is justified because, as a rule,
the dielectric relaxation time, $\tau _{D},$ is much greater than $\tau
_{c}\simeq 1$ns, which is the maximal characteristic time of the
polarization change in our case. Increasing the ionic conductivity of water
by adding, for example, NaCl, one can, according to our estimations,
decrease $\tau _{D}$ down to 0.1ns. In such an electrolyte the mechanism of
SBSL discussed in this paper might be less effective.

\section{The scenario: estimation of the main parameters}

The dynamics of bubble cavitation have been studied in many papers \cite
{review}. Here we will only discuss the short time interval when the bubble
radius $R$ is close to its minimum value $R_{c}$ (Fig.1a). Fig.1a reflects
the most essential features of Fig.4 of Ref.\cite{Barb92} and Fig.12 of Ref.%
\cite{review} where experimental results were presented. Within this
interval the velocity of the bubble surface $v=dR/dt$ reaches its maximum
and reaches zero at the point $R=R_{c}$ (Fig.1b) and the acceleration $%
a=d^{2}R/dt^{2}$ reverses its sign and can achieve huge values (Fig.1c).
During the negative acceleration period ($t\sim t_{1}$, see Fig. 1c) the
gradient of pressure $\nabla p=-\rho a$ is directed from the bubble center
to its periphery, thus according to Eq.\ref{flex} the flexoelectric
depolarizing field has the same direction (Fig.1d). That means that free
electrons that could be generated in the gas or in water near the bubble
surface cannot be accelerated. During the positive acceleration period ($%
t_{1}<t<t_{2}$, see Fig. 1e) the situation is opposite: the arising
flexoelectric field is directed to the bubble center and the generated
electrons can be accelerated by the field and produce a breakdown (Fig. 1e).
According to experimental data \cite{Barb92,Wen97} for a bright SBSL the
characteristic value of the acceleration is about (10$^{11}\div $ $10^{13})$
m/s$^{2}$ during an interval, $\tau _{c}=t_{2}-t_{1}\sim $ 1ns , so the
pressure gradient $\nabla p=-\rho a$ can reach (10$^{14}\div $10$^{16})$N/m$%
^{3}$. Note that the gradient of pressure decays with distance from the
bubble surface . For purpose of estimations one can consider water as
incompressible liquid where $\nabla p=(\nabla p_{s})R^{2}/r^{2}$ where $%
\nabla p_{s}$ is the gradient of pressure at the bubble surface. The same
estimation for the gradient of pressure can be obtained if we take into
account that the pressure is about (10$^{6}\div $10$^{8}$) N/m$^{2}$ at the
final stage of the bubble collapse and the extension of the high pressure
region is about 1$\mu $m \cite{review}.

It follows from Eqs.\ref{flex} and \ref{El} that for $\nabla p$=(10$%
^{14}\div $10$^{16})$N/m$^{3}$ the value of the flexoelectric field, $E_{%
\text{ ,}}$ can reach (10$^{7}\div $ 10$^{9})$V/m. According to Eq.\ref{flex}
the electric field decays just as the gradient of pressure, i.e., the
flexoelectric field in water is given by the formula $E\simeq
E_{s}R^{2}/r^{2}$ where $E_{s}$ is the field at the bubble surface, i.e. the
radial extension of the strong field region is about 1$\mu $m. The field $E_{%
\text{ }}$ can essentially exceed the threshold field of dynamic electric
breakdown of water $E_{th}$ which is about (1$\div $3)$\cdot $10$^{8}$V/m%
\cite{Sachi}. It does not mean, however, that electric breakdown will occur:
a ''lucky electron'' which capable of provoking an avalanche is needed. A
similar situation takes place with the laser breakdown \cite{Sachi}. At the
same time the conduction electron concentration, $n,$ in water at room
temperature is practically zero ($n\sim 10^{-90}$cm$^{-3}$): water could be
considered as a wide gap amorphous semiconductor with $E_{g}\simeq 6.5$eV
\cite{Sachi,Will74,Boyle69,Grand79,Krebs84}.

The breakdown starts only the moment when the strong flexoelectric field is
directed to the bubble center (Fig. 1f) and conduction electrons appear
because of the sharp increase of temperature near the bubble surface . The
latter takes place when the bubble radius $R\simeq R_{s}$ which is close to $%
R_{c}\simeq $(0.5$\div $1)$\mu $. At this moment the flexoelectric field $%
E_{f}$ may be much more greater than $E_{th}$ so the coefficient of
avalanche multiplication of an electron can be practically infinite and in
this case just several electrons are enough to provoke the breakdown and to
screen the field.

%%%%%%%%  Figure goes on a separate page: %%%%%%%%%%
\begin{figure*}
\includegraphics*[height=19cm]{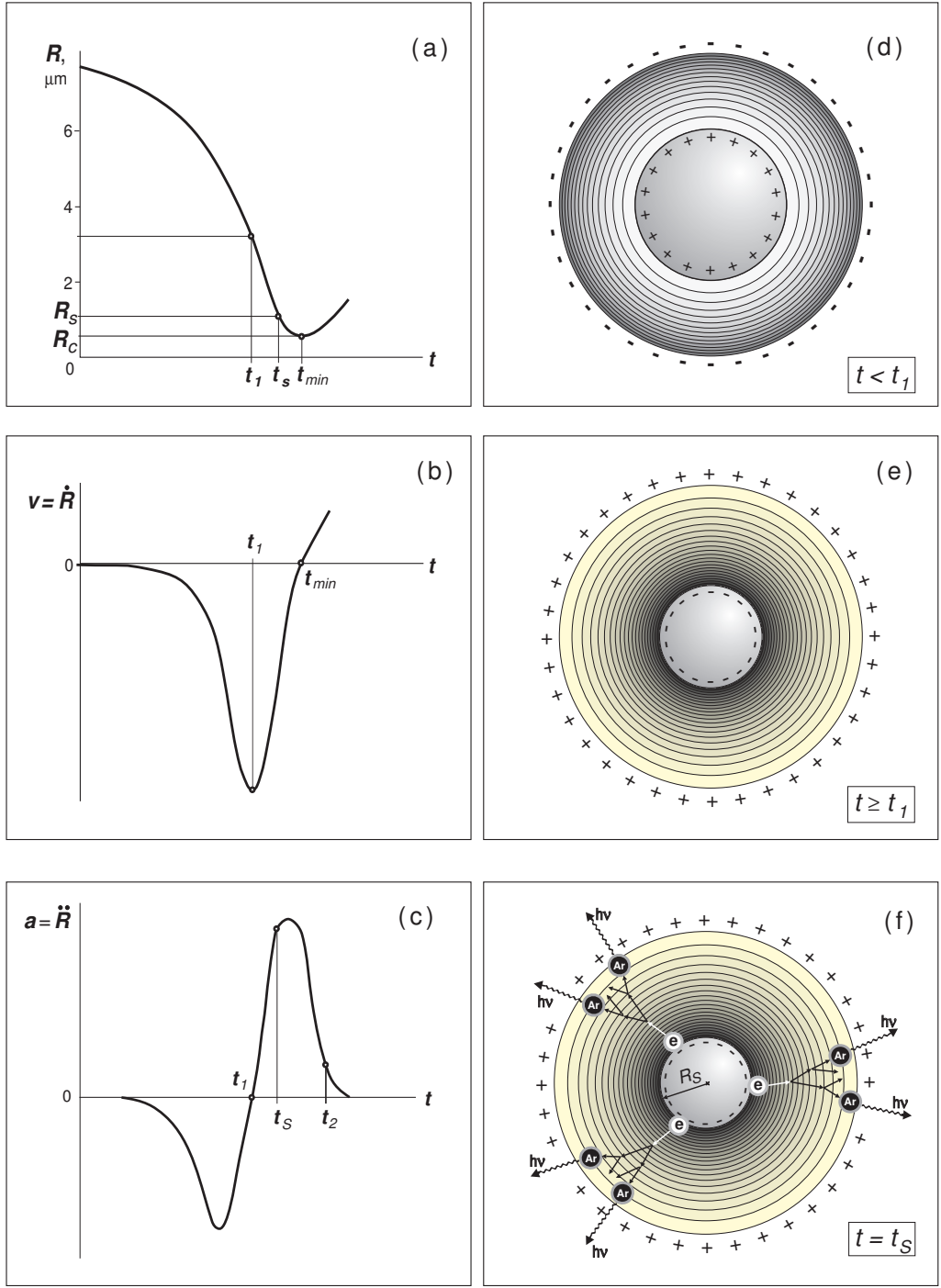}
\caption{Illustration of dynamics of the pulsating bubble and of
formation of the flexoelectric field and the electric breakdown at
the bubble collapse. Time dependence of the bubble radius (a),
velocity (b) and acceleration (c) of the bubble surface. The
flexoelectric polarization at different moments: (d) - at the
beginning of the collapse ($t<t_1)$, (e) - not very close to the
turning point, (f) - in the vicinity of the turning point
($R=R_c).$} \label{fig1}
\end{figure*}
%%%%%%%%%%%%%%%%%%%%%%%%%%%%

Let us show that even relatively moderate temperatures near the bubble
surface (which seem to be generally acceptable for the regimes without the
shock waves) are quite sufficient to provide a sufficient number of
conduction electrons to ensure the breakdown. As in Ref. \cite{Lofs93} we
assume that the gas has the temperature of 7000K. That means that the water
layer with a thickness of the thermal penetration length, $\delta _{T}$, has
a temperature of about 3000K. For $\delta _{T}$ one has:

\begin{equation}
\delta _{T}=(\frac{2\kappa }{\rho c}t)^{1/2}  \label{thermodiff}
\end{equation}
where $\kappa $, the thermal conductivity of water is about 0.4 J/m s K; $c$%
, the specific heat, is about 4$\cdot 10^{3}$ J/m$^{3}$ K; $\rho $, the
density, is 10$^{3}$kg/m$^{3}$. Bearing in mind that the characteristic time
for the last stage of the collapse is $\tau _{c}\sim $1ns \cite{review}, one
finds that the thermal penetration length is $\delta _{T}\sim 0.1\mu $m.
Following \cite{Will74,Sachi,Krebs84} we will consider water as an amorphous
semiconductor with a band gap $E_{g}\simeq 6.5$eV and an effective density
of states $N^{*}$ $\simeq 10^{21}$ for $T\simeq 3000$K. The equilibrium
concentration of conduction electrons in water at $T\geq 3000$K is $%
n=N^{*}\exp (-E_{g}/2kT)\geq 3\cdot 10^{15}$cm$^{-3}$. Taking into account
that the volume of the hot surface layer is approximately $V\simeq
10^{-12}\div 10^{-13}$cm$^{3}$ we find that the equilibrium number of
conduction electrons in the layer is $N>300\div 3000.$ It is important also
to find the time needed for the equilibrium concentration of electrons to be
established. The thermal ionization frequency (see, e.g., \cite{semi}) is

\begin{equation}
\nu _{T}=(N^{*}\sigma v_{T})\exp (-E_{g}/2kT),  \label{thermo}
\end{equation}
where $\sigma $ is the cross-section of the free carrier recombination, $%
v_{T}$ is the thermal velocity of conduction electrons in water. Taking into
account that a reasonable value of $\sigma $ is about 10$^{-15}$cm$^{-2}$
and $v_{T}\sim 10^{7}$cm/s we find from Eq.\ref{thermo} that for $T>3000$K
the transient time to the equilibrium $\tau _{T}=\nu _{T}^{-1}\leq 10^{-8}$%
s. From that follows that during the characteristic time for the last stage
of the bubble collapse $\tau _{c}\sim $1ns the number of conduction
electrons in the layer, $N,$ exceeds $30\div 300$ electrons, which is quite
enough to provoke breakdown.

Note that a similar number of free electrons can be generated by thermal
ionization of the gas. Indeed, the assumed gas temperature (7000K) is about
two times larger and the ionization energy both of the water vapor and Ar,
Kr, Xe is about 12$\div $16eV, i.e. also about two times larger than $E_{g}$%
. Authors of many works (see \cite{review} and the references therein) state
that much higher temperatures can be reached because of shock waves forming
in the bubble gas during the collapse. In principle, one can imagine a
situation where high temperatures are not reached (and free electrons are
not generated) before the shock waves are formed. In this case the shock
wave, when it explodes, will start the electric breakdown.

As a result of breakdown the depolarizing flexoelectric field becomes
screened. The total transmitted charge in the process of the screening is

\begin{equation}
Q_{t}=PS=4\pi R_{s}^{2}\varepsilon _{0}E_{s}  \label{charge}
\end{equation}
where we have taken into account that in our case $D=P+\varepsilon _{0}E=0$,
i.e., $P=-\varepsilon _{0}E$. Assuming $R_{s}\simeq 1\mu $m and $%
E_{s}=(10^{8}\div 10^{9})$ V/m we find that the maximum total number of
transmitted electrons is $N_{t}=Q_{t}/e\simeq (10^{5}\div 10^{6})$.

Note that for considered electric fields the time of the screening
(breakdown), $\tau _{s}$, is determined by the time it takes for the
conduction electrons to cross the region of the strong electric field whose
size is $l\sim R_{s}$. Thus $\tau _{s}=l/v_{d}$ where $v_{d}$ is the
electron drift velocity which in strong electric fields saturates at some
value $v_{d}\simeq 10^{5}$m/s (see Sec.4) . Assuming $l\sim 1\mu $m we find $%
\tau _{s}\sim 10$ps. Note that the value of $\tau _{s}$ is less than the
observed SBSL pulse width \cite{review,Gompf97,Hill98}.

The whole period of positive acceleration, $\tau _{c}$, (see Sec.3) is about
two orders of magnitude larger than $\tau _{b}.$ Therefore, after the
breakdown is finished and the depolarization field is screened the
polarization continues to change, because of change in acceleration. The
electric field arises once more and can exceed the breakdown threshold. As a
result a new breakdown will take place. Such a situation can be repeated
several times. Effectively, it manifests itself in an increase of the pulse
duration which may achieve a fraction of a nanosecond. Therefore, within our
scenario the greater is the pulse duration the greater is its energy.
Analogous interrelation is observed in experiment. \cite
{review,Gompf97,Hill98}.

\section{Energy Distribution of Hot Electrons in the Electric Field and
Spectrum of the Sonoluminescence}

We have already mentioned that the electric fields under consideration are
very strong, inhomogeneous and change over time. However, the characteristic
space and time scales of the field change are correspondingly $l\sim 1\mu $m
and $\tau _{c}\sim 1$ns and they are much more than the relaxation length $%
\lambda $ and relaxation time $\tau _{\varepsilon }$ of the hot electron
energy which are respectively about $(10\div 100)\AA $ and $10^{-13}$s. \cite
{Will74,Sachi,Krebs84}. Therefore one can consider the local electron energy
distribution function using well known results for homogeneous static
electric fields \cite{Wolf,Bafar,Keldysh}.

The considered electric fields are superstrong, i.e. the following condition
is valid:

\begin{equation}
qE\lambda \gg h\omega _{ph}\equiv \varepsilon _{ph},  \label{cond1}
\end{equation}
where $\epsilon _{ph}$ is the characteristic energy of local oscillations in
water which is practically equal to the energy of optical phonons in ice, $%
\epsilon _{ph}\simeq (80\div 100)$meV \cite{Will74,Sachi,Krebs84}. Using $%
\lambda \simeq $ $(10\div 100)\AA $ one sees that the condition \ref{cond1}
is satisfied for fields $E>10^{7}$V/m. The condition \ref{cond1} means that
in these fields an electron acquires, on average, an energy $qE\lambda $
which is much more than $\epsilon _{ph}$. Since in the process of the
acceleration an electron generates many phonons the electron energy
distribution is nearly isotropic in the momentum space. Specifically, in
this case the energy distribution of hot conduction electrons is
approximated with high accuracy, up to energies of electrical breakdown $%
\epsilon =E_{g}$, by the Maxwell function \cite{Wolf,Bafar,Keldysh}:

\begin{equation}
f(\epsilon )\sim \exp (-\frac{\epsilon }{kT_{e}})  \label{distr}
\end{equation}
with an effective electron temperature $T_{e}$ which is determined by the
balance equations for energy and momentum of the hot electrons
\begin{equation}
\frac{d\epsilon }{dt}=eEv_{d}-\frac{\epsilon _{ph}v_{T}}{\lambda }=0,
\label{bal1}
\end{equation}

\begin{equation}
\frac{dp}{dt}=eE-m_{e}v_{d}\frac{v_{T}}{\lambda }=0.  \label{bal2}
\end{equation}
where $v_{T}$ is the effective thermal velocity of electrons. Since for the
Maxwell distribution

\begin{equation}
\frac{3}{2}kT_{e}=\frac{1}{2}m_{e}v_{T}^{2},  \label{vel-temp}
\end{equation}
from Eqs.\ref{bal1}, \ref{bal2} and \ref{vel-temp} it follows that

\begin{equation}  \label{vel-dr}
v_d=\sqrt{\varepsilon _{ph}/m_e}
\end{equation}
and

\begin{equation}
kT_{e}=(eE\lambda )^{2}/3\varepsilon _{ph.}  \label{temp-el}
\end{equation}
From Eqs \ref{vel-temp} and \ref{vel-dr} as well from Eq. \ref{cond1} one
sees that $v_{T}/v_{dr}=\sqrt{kT_{e}/\varepsilon _{ph}}=qE\lambda
/\varepsilon _{ph}>>1.$ This is precisely the condition of validity for the
thermalization of the hot electrons and use of the Maxwell distribution.

Conduction electrons in water form polarons which are conventionally called
hydrated electrons \cite{Boyle69,Krebs84}. However, in the strong electric
field the polaron states decay and the current carriers are the usual
conduction electrons. For purpose of estimation we assume that their
effective mass $m_{e}$ is close to that of free electrons $m_{0}$. Then from
Eq.\ref{vel-dr} it follows that $v_{d}\simeq 10^{5}$m/s for $\epsilon
_{ph}\simeq 100$meV. It has already been mentioned that the electric fields
under consideration are $E\simeq (10^{8}\div 10^{9})$V/m. It follows from Eq.%
\ref{temp-el} that for such fields $kT_{e}\simeq (1\div 10)$eV, i.e., $%
T_{e}\simeq (10^{4}\div 10^{5})$K for $\lambda \simeq (10\div 100)\AA $ .
One sees that the electron temperatures can be several orders higher than
the gas temperature in the bubble. Recall that for our scenario (see Sec.3)
it sufficient that the gas temperature be about 7000K.

Now we will discuss the role of noble gases dissolved in water. An important
feature of an noble gas atom is the existence of metastable states. The life
time of the metastable states can reach several milliseconds when the noble
gas atoms are impurities in solids (see, e.g., \cite{Baum62,Karlov}). Since
the nearest order in water is essentially the same as in solids it would be
natural to expect that the life-time of the metastable states for the noble
gas atoms in water is not less than $1$ns . Hot electrons not only generate
new conduction electrons (the breakdown avalanche) but also excite the noble
gas atoms into metastable states. Below we will estimate from the
experimental data that the characteristic value of $kT_{e}\simeq (2\div 5)$%
eV. However, a considerable number of electrons may have energies about 10eV
and can excite the noble gas atoms.

Note that the energies necessary for excitation to the metastable states, $%
\varepsilon _{m}$, for Xe, Kr, Ar, Ne, He are about 10eV and increase
monotonically from Xe to He. In other words, it is much easier to excite Xe,
Kr and Ar than Ne and He. This could be the reason for the increase in SBSL
intensity in the series He-Xe, which has been observed experimentally \cite
{review,Hills94}.

As we have mentioned above, the life-time of the metastable states of noble
gas atoms in water is expected to be fairly long. Thus, once excited a noble
gas atom can remain in a metastable state for the entire time interval of
positive acceleration and multiple breakdowns, $\tau _{c}\sim 1$ns. The
major part of electrons has the energies $\varepsilon <\varepsilon _{m\text{
,}}E_{g};$they collide with these metastable atoms and transfer them to
higher excited states. The radiation transitions between high-energy excited
states of noble gas atoms govern the SBSL spectrum.

Note that the life-time of the metastable states can exceed the period of
the acoustic wave. In this case there will be an effect of accumulation of
the noble gas atoms in metastable states resulting in the gradual build up
of the SBSL power during several periods of the acoustic wave.

A specific feature of a noble gas is an abundance of excited states with
energies higher than those of metastable states and with high probabilities
of the radiation transitions between them. That is why the optical spectra
of the noble gas atoms contain many lines \cite{Baum62,Karlov}. Because of
the Stark effect in the strong electric field these lines are split.
Besides, in the active region the strong electric field changes at least
several times and the observed radiation spectrum is a superposition of the
spectra of the atoms in different strong electric fields. In other words one
can consider the density of the atomic excited states as a constant.

The probability of a hot electron having an energy $\varepsilon <E_{g}$ is
given by Eq.\ref{distr} and at every collision it transfers the noble gas
atom in a metastable state to a state with energy $\varepsilon $, the
reference point of energy being the energy of the metastable state (for our
estimations we consider only one metastable state). The concentration of
atoms excited during the time of a single electric breakdown, i.e. the
screening time, $\tau _{s},$ to energies within an interval $d\varepsilon $
reads

\begin{equation}
dn_{n}^{**}=n_{n}^{*}(\sigma _{ex}v_{T}n)\tau _{s}\exp (-\frac{\varepsilon }{%
kT_{e}})\cdot \frac{d\varepsilon }{kT_{e}}  \label{numbex}
\end{equation}
where $n_{n}^{*}$ is the number of the noble gas atoms in the metastable
state, $\sigma _{ex}$ is the cross section of the impact excitation of an
atom from the metastable state to a state with energy $\varepsilon $. Atoms
excited to states with the energy $\varepsilon $ generally do not go to the
ground state directly but through intermediate excited states. It reasonable
to assume that with a high probability they radiate phonons with energy $%
h\nu \simeq \varepsilon $. So the spectral density of the SBSL radiation
energy per pulse for unit volume can be written as

\begin{equation}
\widetilde{P}(h\nu )d(h\nu )=h\nu w_{r}n_{n}^{*}(\sigma _{ex}v_{T}n)\tau
_{s}\exp (-\frac{h\nu }{kT_{e}})\frac{d(h\nu )}{kT_{e}}  \label{spect}
\end{equation}
where $w_{r}$ is the probability of the spontaneous radiation transition.
Taking into account that $w_{r}=\frac{4(2\pi )^{4}\nu ^{3}}{3c^{3}h}D^{2}$
where $D$ is modulus of matrix element of the dipole moment of the
transition \cite{Bete} we find

\begin{equation}
\widetilde{P}(h\nu )d(h\nu )=\frac{4(2\pi )^{4}D^{2}}{3c^{3}}
n_{n}^{*}(\sigma _{ex}v_{T}n)\tau _{s}\nu ^{4}\exp (-\frac{h\nu
}{kT_{e}}) \frac{d(h\nu )}{kT_{e}}. \label{spectrn}
\end{equation}

Recall that $T_{e}$ in Eq. \ref{spectrn} is a function of the coordenates
because $T_{e}\sim E^{2}$ (see Eq.\ref{temp-el} ). Putting, as before, $%
E=f\nabla p\simeq E_{s}R_{s}^{2}/r^{2}$ and integrating approximately Eq.\ref
{spectrn} we find the spectral power of SBSL for a single breakdown is

\begin{equation}
P=\int \widetilde{P}(h\nu ,r)dV=\frac{4(2\pi )^{4}D^{2}}{3c^{3}h}%
n_{n}^{*}(\sigma _{ex}v_{T}N_{t})\tau _{s}\nu ^{3}\exp (-\frac{h\nu }{kT_{es}%
})  \label{SBSL}
\end{equation}
where $kT_{es}=(eE_{s}\lambda )^{2}/3\varepsilon _{ph\text{ }}$and $N_{t}$
is, practically, the total number of electrons participating in the
breakdown. Note that the approximative Eq.\ref{SBSL} is not sensitive to the
form of decay of the field, it is important only that $E$ decays more
steeply than $r^{-1}$, i.e. it is not important that for $\nabla p$ we use
an expression for an incompressible liquid.

The observed spectra are cut off in the shortwave region because of the
absorption of water. This can be taken into account by multiplying Eq.\ref
{SBSL} by exp(-$\alpha (h\upsilon )L$) where $L$ is the size of the acoustic
resonator ($L\simeq 2.5$cm \cite{review}). Because of the Urbach absorption
tails \cite{Will74,Grand79} the radiation maximum is located considerably
below than $h\nu \simeq E_{g}\simeq 6.5eV$.

It should be emphasized that the spectrum given by Eq.\ref{SBSL} resembles
the black-body spectrum but temperature is given here by the effective
temperature $T_{es}$ of hot electrons near the bubble surface. The value of $%
T_{es}$ may be much higher than the bubble gas temperature. Experimentally
observed spectra can be fitted, in the wavelength interval (200$\div 700)$
nm, to the black-body ones with temperatures (2$\div $5)10$^{4}$K \cite
{review,Hill92}. According to Eq.\ref{temp-el} such electron temperatures
are reached at electric fields $E_{s}\simeq (2\div 10)10^{8}$ V/m for values
of $\varepsilon _{ph}\simeq 0.1$eV and $\lambda \simeq (10\div 100)$ \AA\
characteristic for water \cite{Sachi}. As follows from Eqs.\ref{flex} and $%
\nabla p=-\rho a$ such fields are realized at accelerations $a\simeq
(10^{13}\div 10^{14})$ m/s$^{2}.$ Similar accelerations are reported in \cite
{review,Wen97}.

Integrating Eq.\ref{SBSL} multiplied by exp(-$\alpha (h\upsilon )L$) we find
an estimate of the radiation energy for a single breakdown

\begin{equation}
W_{r}=\int Pw_{r}^{-1}exp(-\alpha (h\upsilon )L)d(h\nu )\simeq (\sigma
_{ex}v_{T}N_{t})\tau _{s}n_{n}^{*}h\overline{\nu }\simeq \frac{v_{T}}{v_{d}}%
(\sigma _{ex}R_{s}n_{n}^{*})N_{t}h\overline{\nu }  \label{power}
\end{equation}
where $h\overline{\nu }$ is the characteristic photon energy which is close
to the energy of the maximum of the observed spectrum, which is about ($%
5\div 6)eV$ in the case of strong electric field of our interest. Recall
that in Eq.\ref{power} $\tau _{s}\simeq R_{s}/v_{d}$ is the screening
(breakdown) time and $N_{t}\simeq 4\pi R_{s}^{2}\varepsilon _{0}E_{s}/e$ is
the number of transmitted electrons in the screening process (Sec. 3) and $%
n_{n}^{*}$ is the concentration of noble gas atoms in a metastable state
near the bubble surface.

As we have mentioned before it was shown in \cite{LohsePRL97,LohseJCP97}
that because of dissolved gas diffusion and chemical reactions an
accumulation of noble gas in the bubble takes place while the bubble
pulsates and in stationary conditions the gas in the bubble consists almost
entirely of noble gas. Therefore one can assume that the concentration of
the noble gas atoms, $n_{n},$near the bubble surface is close to its
saturation value. Note that this value has to be estimated for atmospheric
pressure and ambient temperature rather than for the high pressures and
temperatures existing at the last stage of the collapse over a very short
time. This is why we assume that $n_{n}\simeq $ (10$^{18}-10^{19}$) cm$^{-3}$%
. Now we will argue that the value of $n_{n}^{*}$ can be only one order of
magnitude less than $n_{n}$. Indeed, for $kT_{e}\simeq (3\div 5)$ eV
obtained above by fitting the SBSL spectrum to the black-body spectrum, the
number of electrons having an energy higher than 10eV, $N_{h}$ may be about 1%
$0^{-3}\div 10^{-2}$ of the total electron number $N_{t}\simeq 10^{6}$ i.e.,
$N_{h}$ is about 10$^{3}\div 10^{4}$. Such a ''superhot'' electron excites a
noble gas atom to the metastable state over the time $\tau _{ex}=(\sigma
_{ex}v_{T}n_{n})^{-1}\simeq 10^{-12}$s for $\sigma _{ex}\simeq 10^{-15}$cm$%
^{2},$ $v_{T}\simeq 10^{6}$ m/s and $n_{n}\simeq $(10$^{18}\div 10^{19}$) cm$%
^{-3}.$ Therefore a ''superhot'' electron during its participation in
screening excites $\tau _{s}/\tau _{ex}$ noble gas atoms and the total
number of the excited atoms is $N_{h}\tau _{s}/\tau _{ex}\simeq 10^{4}\div
10^{5}$ if one takes into account that $\tau _{s}\simeq 10\tau _{ex}$ $%
\simeq 10^{-11}$s. The total number of the noble gas atoms in the region of
the strong field ($\sim $1$\mu $m$^{3})$ is about $10^{6}\div 10^{7}$, i.e.
the percentage of the noble gas atoms excited to the metastable state during
a single breakdown can be about 10$\%$. Since the life time of the
metastable state is much more than $\tau _{c}$ and during this time interval
several breakdowns can take place (Sec.3), one may expect that the number of
the noble gas atoms excited to the metastable state during the interval $%
\tau _{c}$ is only one order of magnitude less than the total number of
these atoms i.e. $n_{n}^{*}$ $\sim $ 10$^{18}$ cm$^{-3}$. If the life time
of the metastable states exceeds the acoustic period ($\sim 30\mu $s) an
accumulation of noble gas atoms in the metastable states can also occur
during several acoustic periods.

Now we can estimate the total radiation energy. We have mentioned above that
the maximum value of $N_{t}\simeq 10^{6}$, $R_{s}\simeq $ 1$\mu $m, $\sigma
_{ex}\simeq 10^{-15}$cm$^{2}$ and $v_{T}>>v_{d}$. Taking into account that
the positive acceleration period $\tau _{c}\sim 1$ns and the breakdown time $%
\tau _{s}$ is about 0.01ns one can assume that the number of breakdowns is
not less than 10. Thus one finds from Eq.\ref{power} that the maximum photon
number in the SBSL pulse can be more than 10$^{7}$. This value corresponds
to the maximum photon number observed experimentally \cite
{review,Hill92,Barb94}. Note that the radiation energy is a small part of
the total energy of the flexoelectric field (see Ref\cite{GLO}).

\section{Conclusions}

Let us emphasize once more that within the mechanism of SBSL considered the
main processes occur in water near the bubble surface unlike the most widely
discussed mechanisms of SBSL \cite{review}. Within the framework of this
mechanism many experimental data about the SBSL can be explained quite
naturally. Of course, since the parameters of the collapsing bubble are not
reliably known when the bubble radius is close to its minimum value we can
present no more than order-of-magnitude estimations.

Let us summarize the main results.

(i). The minimum duration of the SBSL flash is determined by the single
breakdown (screening) time, $\tau _{s}\simeq 10$ps. A larger time is
possible because of the multiplicity of breakdowns. The maximum duration is
limited by $\tau _{c}\sim 1$ns. The longer is the duration the greater is
the energy of the flash. This is in agreement with experiments where it was
found that the pulse duration changes from 30ps to 400ps when its energy
increases \cite{review,Gompf97,Hill98}.

(ii). According to our estimations the maximum energy of the flash
corresponds to $10^{7}\div 10^{8}$ photons with energy $(5\div 6)$eV, which
also agrees with the experiment \cite{review,Hill92,Barb94}.

(iii). Within our scenario the noble gas atoms play an important role. They
do not reduce so much the breakdown threshold as they govern the radiation
process. Abundance of optical transitions in these atoms and inhomogeneous
broadening because of the Stark effect explain the practically continuous
character of the SBSL spectrum.

(iv). In agreement with the experiment \cite{review,Hill92} the theoretical
spectrum of the SBSL resembles the black-body spectrum but the temperature
is given here by the effective temperature of the hot electrons which can be
about several eV, what corresponds to the observed apparent radiation
temperature \cite{review,Hill92}. At the same time the gas temperature is
not directly related to the radiation temperature and can be considerably
less than 1eV.

(v). The increase of influence on SBSL intensity in the noble gas series
He-Xe observed in the experiment \cite{review} is connected with the
decrease, in this series, of the energy of the lowest metastable state.

(vi). The pulse width does not depend on the spectral range of the radiated
photons. This also agrees with experiment \cite{Gompf97}.

(vii). The effect of synchronization of the light pulses observed
experimentally \cite{review,synchr} is explained.

Note also that within our scenario of SBSL the influence of magnetic fields
on SBSL is fairly weak. Their influence becomes appreciable when they are
high enough to hamper the heating of electrons\cite{Bass}. This might be the
reason for the decrease of the SBSL intensity in strong magnetic fields
which has been observed experimentally\cite{MF}.

It seems that the considered mechanism of SBSL is especially effective for
water because of a lucky coincidence of several conditions:

(i) the strong geometric asymmetry of the water molecule, which accompanies
its electric asymmetry, provides the needed sign of the flexoelectric
coefficient. This sign is such that the electric field has the ''correct
direction'' (accelerates electron) over the same (very short) period in
which electrons are generated because of the sharp increase of the bubble
gas temperature. The temporal coincidence of these two cicumstances is a
possible reason of the effect of synchronization of the light pulses
observed experimentally \cite{review,synchr}.

(ii) water is a semiconductor with a relatively narrow band gap ($E_{g}=6.5$%
eV) and sufficienly wide conduction band, what is necessary for the heating
of the electrons in strong electric field.

(iii) high solubility of noble gases in water what makes possible high
concentrations of noble gas in water near the bubble filled by the noble gas
accumulated there in the process of the bubble pulsations.

Fulfillment of these conditions gives the answer to the question why water
is the friendliest liquid for SBSL \cite{review}.

Some experimentally found characteristics of SBSL have not been explained in
this paper, in particular the dependence of the energy of the pulse upon the
partial pressure of noble gases and water temperature \cite{review}. These
dependences may be not necessarily a manifestation of the radiation
processes studied in this paper but they could be a consequence of a set of
various factors. They may be determined by the hydrodynamics of the
pulsating bubble, the solubility of noble gases in water and also by the
temperature dependence of the flexoelectric coefficient of water which still
remains, unfortunately, unmeasured.

Although the main features of SBSL seem to be explainable within our
scenario, these explanations are qualitative rather than quantitative. We
are still a long way from a quantitative theory at present. One of the main
aims of the paper is to stimulate some experiments that could either support
or discard the proposed mechanis of SBSL.

(i) Measurements of flexoelectric coeficients of water.

(ii) A detailed study of radiation spectra at electric breakdown of water
and ice with different concentrations of noble gases.

(iii) Study of influence of water conductivity on the SBSL intensity.

After performing these experiments it makes sense to develop the theory
further. In particular, it will be necessary to study in more detail the
kinetics of the screening process and of the excitation of noble gas atoms
while taking into account the time variation of the distribution function of
the hot electrons. This problem should be considered together with that of
the determination of the spatial distribution of excited noble gas atoms
near the bubble surface. Diffusion and relaxation of the excited atoms
should be considered together, of course, with the processes that lead to
noble gas accumulation in the bubble \cite{LohsePRL97,LohseJCP97}.

In our estimations we assumed that flexoelectric coefficient of water has
its ''natural'' value. This was sufficient for the theoretical estimations
to be in agreement with the experimental data. In fact, because of the
above-mentioned strong geometric asymmetry of the water molecule, the
coefficient could be even larger. If this is the case some new phenomena
will take place including influence of the flexoelectric effect on
hydrodynamics of the pulsating bubble and X-ray radiation induced by
electrons of very high energy.

We are grateful to V.Kholodnov for useful discussions. NG and VVO thank EU
ESPRIT, the Spanish CSIC and NATO for Linkage Grant Ref. OUTRLG 970308.
Also, APL thanks NATO for Linkage grant HTECHLG 971213.

%%%%% References: %%%%%

%%%%%% Figure caption %%%%%
%\begin{figure*}[tbp]
%\caption{Illustration of dynamics of the pulsating bubble and of formation
%of the flexoelectric field and the electric breakdown at the bubble
%collapse. Time dependence of the bubble radius (a), velocity (b) and
%acceleration (c) of the bubble surface. The flexoelectric polarization at
%different moments: (d) - at the beginning of the collapse ($t<t_1)$, (e) -
%not very close to the turning point, (f) - in the vicinity of the turning
%point ($R=R_c).$}
%\label{fig1}
%\end{figure*}
%
%%%%%%%%%  Figure goes on a separate page: %%%%%%%%%%
%\newpage
%\begin{figure*}
%\includegraphics*{sbslfig.ps}
%\end{figure*}
%%%%%%%%%%%%%%%%%%%%%%%%%%%%%


\begin{references}
\bibitem{LC}  The term ''flexoelectric effect'' is also used in the liquid
crystals physics but it has a different meaning.
\end{references}
\end{document}